%% file: main.tex
\documentclass[conference]{IEEEtran}
\IEEEoverridecommandlockouts
\usepackage{cite}
\usepackage{amsmath,amssymb,amsfonts}
\usepackage{graphicx}
\usepackage{textcomp}
\usepackage{xcolor}
\usepackage[a4paper, total={184mm,239mm}]{geometry}
\def\BibTeX{{\rm B\kern-.05em{\sc i\kern-.025em b}\kern-.08em
    T\kern-.1667em\lower.7ex\hbox{E}\kern-.125emX}}

\usepackage{amsmath,amssymb,amsfonts}
\usepackage{graphicx}
\usepackage{algorithm}
\usepackage{algpseudocode}
\usepackage{threeparttable}
\usepackage{graphics}
\usepackage{epsfig}
\usepackage{textcomp}
\usepackage{xcolor}
\usepackage{pifont}
\usepackage{amssymb}
\usepackage{booktabs}
\usepackage{multirow}
\usepackage{makecell}
\usepackage{hyperref}
\usepackage{mathtools}

\newcommand{\chao}[1]{\textcolor{black}{#1}}
\newcommand{\qi}[1]{\textcolor{black}{#1}}
\newcommand{\zong}[1]{\textcolor{black}{#1}}
\newcommand{\qititle}[1]{\textcolor{black}{#1}}
\newcommand{\chaoRtwo}[1]{\textcolor{black}{#1}}
\newcommand{\qicamera}[1]{\textcolor{black}{#1}}

\begin{document}

\title{A Scheduling Framework for Efficient \\ MoE Inference on Edge GPU-NDP Systems}



\author{
     Qi Wu\textsuperscript{1}, 
     Chao Fang\textsuperscript{1,$\dagger$}, 
     Jiayuan Chen\textsuperscript{2}, 
     Ye Lin\textsuperscript{1}, 
     Yueqi Zhang\textsuperscript{1}, 
     Yichuan Bai\textsuperscript{1},
     Yuan Du\textsuperscript{1}, 
     Li Du\textsuperscript{1} \\
	\IEEEauthorblockA{
		\textsuperscript{1}School of Electronic Science and Engineering, Nanjing University, China~~~\textsuperscript{2}China Mobile Research Institute, China \\
        \small Email:
		\{qiwu, fantasysee, yelin, zhangyueqi, yicbai\}@smail.nju.edu.cn, chenjiayuan@chinamobile.com, \{yuandu, ldu\}@nju.edu.cn
    }
}

\maketitle

\input{Sections/0-abstract.tex}
\input{Sections/1-intro.tex}
\input{Sections/2-background.tex}
\input{Sections/3-framework.tex}

\input{Sections/4-experimental.tex}
\input{Sections/5-conclusion.tex}

\bibliographystyle{IEEEtran}
\bibliography{ref}

\end{document}

%% file: Sections/0-abstract.tex
\begin{abstract}

Mixture-of-Experts (MoE) models facilitate edge deployment by decoupling model capacity from active computation, yet their large memory footprint drives the need for GPU systems with near-data processing (NDP) capabilities that offload experts to dedicated processing units.
However, deploying MoE models on such edge-based GPU-NDP systems faces three critical challenges: 1) severe load imbalance across NDP units due to non-uniform expert selection and expert parallelism, 2) insufficient GPU utilization during expert computation 
\chao{within}
NDP units, and 3) extensive data pre-profiling necessitated by unpredictable expert activation patterns for pre-fetching.
To address these challenges, this paper proposes an efficient inference framework featuring three key optimizations. 
\chao{First, the underexplored tensor parallelism in MoE inference is exploited to partition and compute large expert parameters across multiple NDP units simultaneously towards edge low-batch scenarios.}
Second, a load-balancing-aware scheduling algorithm distributes expert computations across NDP units and GPU to maximize resource utilization. 
Third, a dataset-free pre-fetching strategy proactively loads frequently accessed experts to minimize activation delays.
Experimental results show that our framework enables GPU-NDP systems to achieve \qicamera{2.41$\times$ on average and} up to 2.56$\times$ speedup in end-to-end latency compared to state-of-the-art approaches, significantly enhancing MoE inference efficiency in resource-constrained environments.
\end{abstract}


\let\thefootnote\relax\footnotetext{$^\dagger$Corresponding author. This work was funded in part by the National Key Research and Development Program of China under Grant 2022YFB4400900, in part by the Strategic Industries and Key Technologies Project of Jiangsu Province under Grant BE2023020-3, in part by the Basic Research Program of Jiangsu Province under Grant BK20243042, and in part by the Nanjing University-China Mobile Communications Group Co., ltd. Joint Institute.}

%% file: Sections/1-intro.tex
\section{Introduction} \label{sec:intro}

\begin{figure} [t]
    \centering
    \includegraphics[width=1\linewidth]{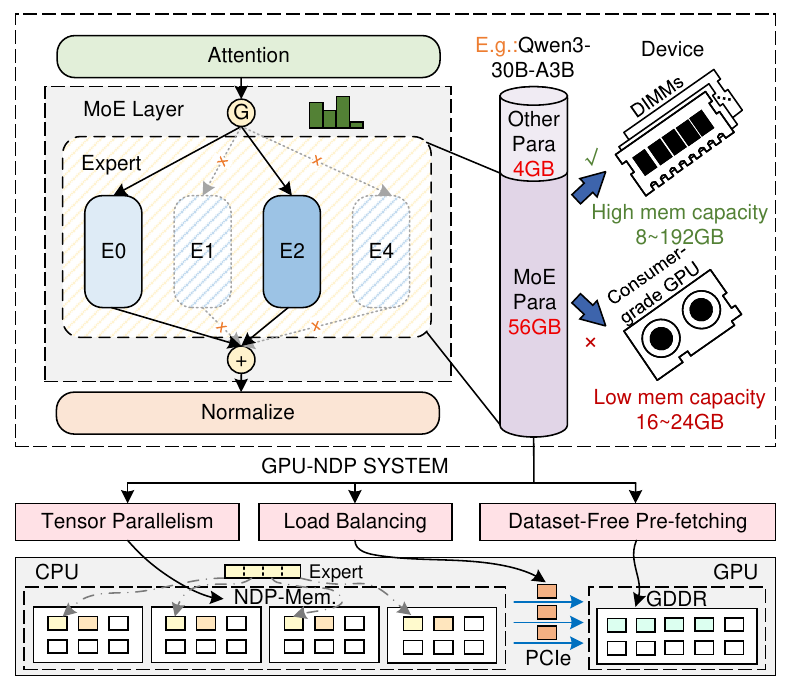}
    \vspace{-18pt}
    \caption{\qititle{The rising memory gap between consumer-grade GPU VRAM and large MoE models motivates GPU-NDP systems as a promising cost-effective edge solution. Our scheduling framework enables efficient MoE inference on such systems via tensor parallelism, load balancing, and dataset-free pre-fetching.}}
    \label{fig:imbalance}
    \qicamera{\vspace{-10pt}}
\end{figure}

\chao{For transformer-based~\cite{vaswani2017attention} large language models (LLMs), the Mixture-of-Experts (MoE) model has recently emerged as an efficient architectural paradigm that enables massive model scaling through sparse computation.}
As shown in Fig.~\ref{fig:imbalance}, the core mechanism of MoEs involves sparse activation in the feed-forward network (FFN) stage, where only a selected subset of parameters, termed "experts," is engaged for each token \cite{jacobs1991adaptive, Shazeer2017Outrageously}, differentiating it from dense models \cite{achiam2023gpt, touvron2023llama, zhang2022opt, fang2025anda} that activate all parameters.
\chao{However, this efficiency comes at substantial memory costs,
as 
expert parameters typically ranging from tens to hundreds of billions~\cite{cai2025survey}, 
far exceeding consumer-grade GPU capacity.}
For instance, RTX 5080, a modern high-end consumer-grade GPU, offers 16GB VRAM \cite{nvidiaRTX5080}, 
\zong{which is insufficient to host many large MoE models, e.g., Qwen3-30B-A3B \cite{yang2025qwen3} whose parameters can occupy more than \qi{60GB}.}
\zong{This memory capacity gap in Fig.~\ref{fig:imbalance}, presents a critical challenge for edge LLM deployment with limited resource budget.}

To address this challenge \chao{of memory capacity}, current research follows mainly three different technical paths.
\qi{Two of them are CPU‑based solutions.}
\chao{The first one leverages host system memory, i.e., CPU DRAM as an extended GPU memory pool, offloading inactive expert parameters and transferring them back when needed \cite{aminabadi2022deepspeed, fang2025klotski, eliseev2023fast}.}
\chao{The second one incorporates heterogeneous computing, where both parameters and computations are partially delegated to the CPU to reduce transmission overhead \cite{zhang2025daop, cao2025moe, kamahori2025fiddler}.}
\chao{Though these approaches could mitigate transmission latency through static pre-fetching\cite{fang2025klotski} or multi-batch strategies\cite{cao2025moe}, they remain fundamentally constrained by memory bandwidth and computational capability of CPU. 
Moreover, static pre-fetching requires extensive pre-runs with calibration data to generate expert usage patterns, limiting its adaptability to edge dynamic scenarios.}



\zong{The third category explores near-data processing (NDP) systems by situating computation proximate to data storage.\cite{kim2024monde, yun2024duplex}.}
\chaoRtwo{These systems can address memory capacity constraints and substantially reduce data transfer latency over the PCIe bus by integrating computation capabilities through pluggable dual in-line memory modules (DIMMs) or modified graphics double data rate (GDDR) into GPU memory systems.}
\chaoRtwo{However, current GDDR-based approaches incur substantial implementation costs~\cite{yun2024duplex}, making them less suitable for cost-effective edge deployments.}
\chaoRtwo{More critically, existing scheduling strategies for GPU-NDP systems suffer from several limitations.}
\chaoRtwo{First, expert parallelism implementations across multi-node systems~\cite{kim2024monde, yun2024duplex} lead to inefficient expert distribution during single-batch inference. Furthermore, heterogeneous scheduling schemes like Duplex~\cite{yun2024duplex} fail to achieve balanced computation between GPU and NDP devices due to consistent arithmetic intensity across experts during the decode stage, while MoNDE~\cite{kim2024monde}, despite introducing load-balancing strategies, relies on periodic adjustments based on historical profiles, resulting in sub-optimal efficiency for dynamic edge workloads. Additionally, the reliance on calibration-based expert prefetching strategies~\cite{kim2024monde} becomes inaccurate in edge inference scenarios where input distributions differ from calibration datasets, leading to reduced efficiency.}
\zong{To address the above limitations,}
this paper proposes a novel scheduling framework for accelerating MoE inference
on edge GPU-NDP systems\qicamera{, specifically targeting localized environments such as personal workstations}, as shown in Fig.~\ref{fig:imbalance}.
\qicamera{The framework is designed for edge deployments with NDP-enabled DIMM configurations, which are preferred over HBM-PIM due to their superior cost-effectiveness and reduced integration complexity. And tensor parallelism is employed to partition expert parameters across devices.}
\chao{Unlike existing approaches that rely on dataset-specific expert activation heatmaps, we dynamically observe expert activation patterns during initial prompt processing phase, i.e., \textit{prefill}, and proactively pre-fetch experts predicted to be active in the subsequent token generation phase, i.e., \textit{decode}, to the GPU.}
\chao{Additionally, when no activated expert resides on the GPU, our load-balancing mechanism transmits only partial expert parameters back to GPU, enabling simultaneous utilization of both GPU and NDP-DIMM resources for computation.}
Our key contributions are summarized as follows. 

\begin{enumerate}

\item A scheduling framework \chao{is proposed} for efficient MoE inference on \chao{GPU-NDP}
systems 
that integrates a consumer-grade GPU and \chao{cost-effective} NDP-DIMMs.
\chao{It is specifically optimized for single-batch inference scenarios common in edge deployment.}
\item A \chao{dynamic} hybrid scheduling strategy is proposed to enhance the efficiency of MoE inference, uniquely combining tensor parallelism with load balancing mechanisms and dataset-free 
expert pre-fetching.
\item \chao{Comprehensive evaluation across four popular MoE models is conducted, 
showing up to 2.56× speedup in end-to-end latency compared with the state-of-the-art (SOTA) MoNDE~\cite{kim2024monde} and demonstrating the effectiveness of our approach for edge MoE inference.}
\end{enumerate}

%% file: Sections/2-background.tex
\section{Background and Related Work}
\label{sec:background}

\subsection{Demand for Edge MoE Deployment}
\qi{Local deployment of MoE models is driven by requirements for offline availability, strengthened data privacy, and user-specific personalization without cloud dependency \cite{ji2024co, hadish2024language, huang2024precision, liu2024spark}.}
\chao{However, the substantial parameter footprint of modern MoE models, such as Qwen3-30B-A3B~\cite{yang2025qwen3} which exceeds 60GB, significantly impedes edge deployment under tight memory and I/O budgets.}

\chao{Empirical evidence reveals that expert activation follows highly uneven patterns: only a small fraction of experts are frequently activated while the majority remain idle for most tokens.}
\chao{Prior systems~\cite{cao2025moe, fang2025klotski} exploit this sparsity through expert pre-fetching and inter-batch weight reuse to amortize loading overheads across large batches for efficient inference.}
\chao{However, edge settings typically involve single-user interaction with a batch size of 1~\cite{cai2023medusa}, eliminating inter-batch amortization opportunities.}
\chao{Moreover, traditional pre-fetching~\cite{fang2025klotski} relies on offline profiling with calibration datasets to determine expert usage patterns, leading to suboptimal efficiency in dynamic edge scenarios.}
\chao{This mismatch between calibration data and actual workloads often results in unused expert transfers during single-batch inference.}
\chao{Consequently, weight traffic between processors and memory becomes the dominant bottleneck in low-batch edge inference. This motivates exploring NDP approaches that bring computation closer to where expert weights reside, rather than continuously transferring large weight matrices.}


\begin{figure} [t]
    \centering
    \includegraphics[width=1\linewidth]{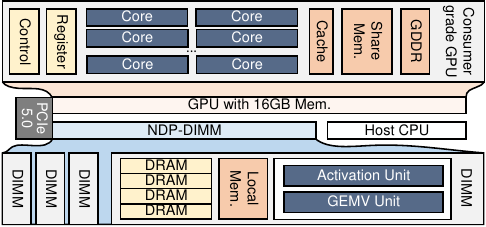}
    \vspace{-18pt}
    \caption{Overview of GPU-NDP DIMM System.}
    \label{fig:system}
    \vspace{-10pt}
\end{figure}

\subsection{Near-Data Processing for MoE Model Acceleration}
\chao{NDP methods~\cite{kim2024monde, yun2024duplex, ke2020recnmp, kwon2019tensordimm, park2024attacc, wu2025pimoe, liu2025make, devaux2019true, he2025lp, jiang2025NDPage, liang2025HyQA, zhang2025Near, zhao2024pim, chen2023bramac, kim2022overview} address the data movement bottleneck by placing computation close to or within memory arrays.}
\chao{In MoE inference, this approach is particularly effective since traffic is dominated by expert weights rather than activations.}
\chao{For example, in single-batch scenarios, the expert weights activated per layer are 672 MB in Mixtral-8x7B~\cite{jiang2024mixtral}, while the activations are only 0.0078 MB. By executing expert computation where weights reside, NDP allows only lightweight activations to be transported.}

\qi{Currently, NDP platforms span HBM-PIM solutions and commodity-module designs based on DIMM or LPDDR devices. 
HBM-PIM methods show strong performance on suitable workloads \cite{park2024attacc, wu2025pimoe}, but \zong{its} reliance on specialized, costly stacked memory limits adoption in cost-sensitive, locally deployed systems. Instead, DIMM-/LPDDR-based NDP leverages standard modules and offers a more economical, scalable path for edge deployments \cite{he2025lp,kim2024monde}.}
\chao{However, existing commodity solutions have significant limitations for edge MoE inference.}
\qi{\zong{LP-Spec~\cite{he2025lp}} employs LPDDR-NDP with NPU to accelerate dense models via dynamic scheduling that coordinates the computation of the attention and FFN.
However, its scheduling is triggered only during token pruning and does not encompass MoE models. 
By contrast, MoE models are explicitly targeted in MoNDE~\cite{kim2024monde}, where expert computations are offloaded to LPDDR-NDP units while retaining other computations on a GPU. It primarily considers a single large-memory NDP device and scales poorly to multi-NDP configurations under single batch, where inter-device workload imbalance significantly lower utilization.}
\chao{These limitations highlight the need for more effective scheduling strategies tailored to multi-NDP edge deployment scenarios, which motivates our proposed framework.}

%% file: Sections/3-framework.tex
\section{The Proposed Framework} \label{sec:framework}


\subsection{Framework overview}
\begin{figure} [t]
    \centering
    \includegraphics[width=1\linewidth]{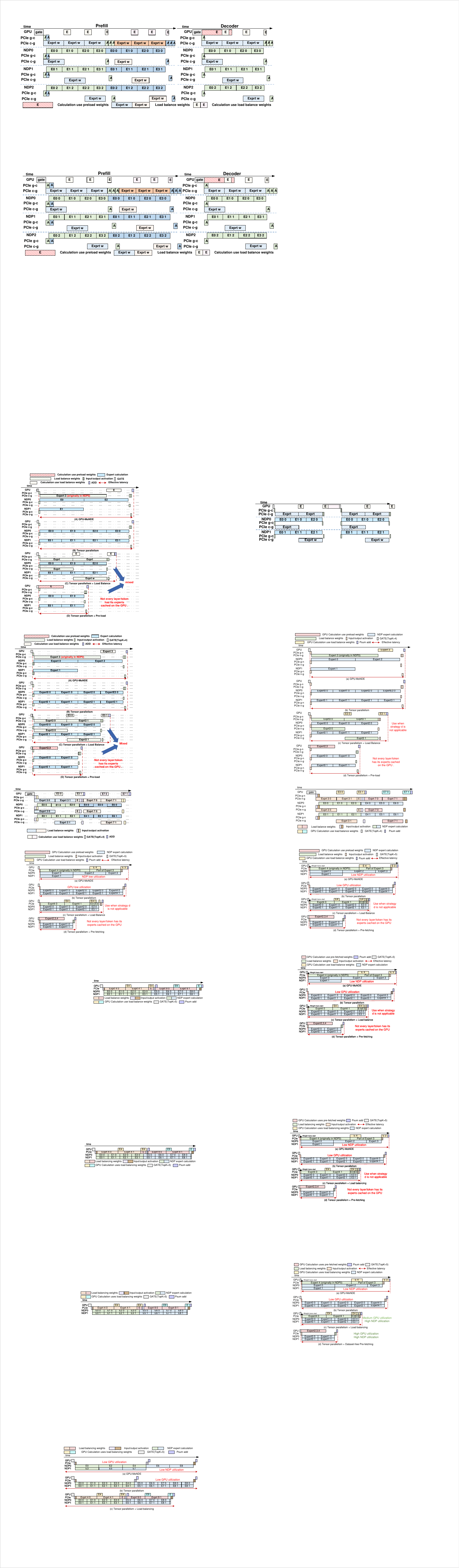}
    \vspace{-18pt}
    \caption{Comparison between MoE workflow scheduling in the decode stage. \qititle{The proposed scheduling framework is based on (b) tensor parallelism and supplemented by (c) load balancing and (d) dataset-free pre-fetching.}}
    \label{fig:decode_strategy}
    \qicamera{\vspace{-10pt}}
\end{figure}

\chao{Fig.~\ref{fig:system} presents the overview of the proposed system framework, which is based on a single consumer-grade GPU and multiple NDP-DIMMs.}
\chao{Our framework} 
adopts 
the central buffer-based NDP-DIMM architecture~\cite{liu2025make, yun2024duplex, cong2017aim, ke2020recnmp}, where an integrated NDP core on each DIMM processes locally stored data.
\chao{Under this GPU-NDP DIMM system, the computational workload is distributed across different processing units.}
\chao{The dominant MoE computations in FFN layers are delegated to DIMM-based NDP units, while all remaining components, including the attention mechanism, are processed on the GPU. The host CPU is reserved for task scheduling and coordination.}
While our analysis 
is grounded in 
this specific hardware configuration, the framework is broadly applicable to the other GPU-NDP systems.
\chao{To optimize MoE inference, we propose three collaborative innovations: (1) a tensor parallelization strategy that distributes expert weights efficiently across NDP units; 
(2) a load balancing scheduling strategy that adapts to token routing patterns; 
(3) a dataset-free pre-fetching strategy that eliminates prior data profiling.}
\chao{Fig.~\ref{fig:decode_strategy} shows these strategies and their incremental benefits over the SOTA MoNDE~\cite{kim2024monde}.}

\subsection{Tensor parallelism}
\chao{The conventional expert parallelism used by MoNDE~\cite{kim2024monde} allocates entire experts to distinct NDPs. However, the dynamic routing mechanism inherent in MoE models does not guarantee uniform expert activation across NDPs, resulting in suboptimal resource utilization as demonstrated in Fig.~\ref{fig:decode_strategy}a. 
\qi{In this case, four activated experts are deployed in NDP0 whereas only one resides in NDP1, rendering NDP0 the critical path. Although partial workloads have been offloaded to the GPU, the critical path remains lengthy.}}

\begin{figure} [t]
    \centering
    \includegraphics[width=1\linewidth]{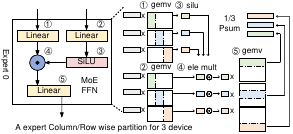}
    \vspace{-18pt}
    \caption{The computing process at the MoE layer with our introduced tensor parallelism. The left side is the common structure of the MoE Expert FFN, and the right side is the deployment of one expert to three computing devices.}
    \label{fig:tp}
\end{figure}

\begin{figure} [t]
    \centering
    \includegraphics[width=1\linewidth]{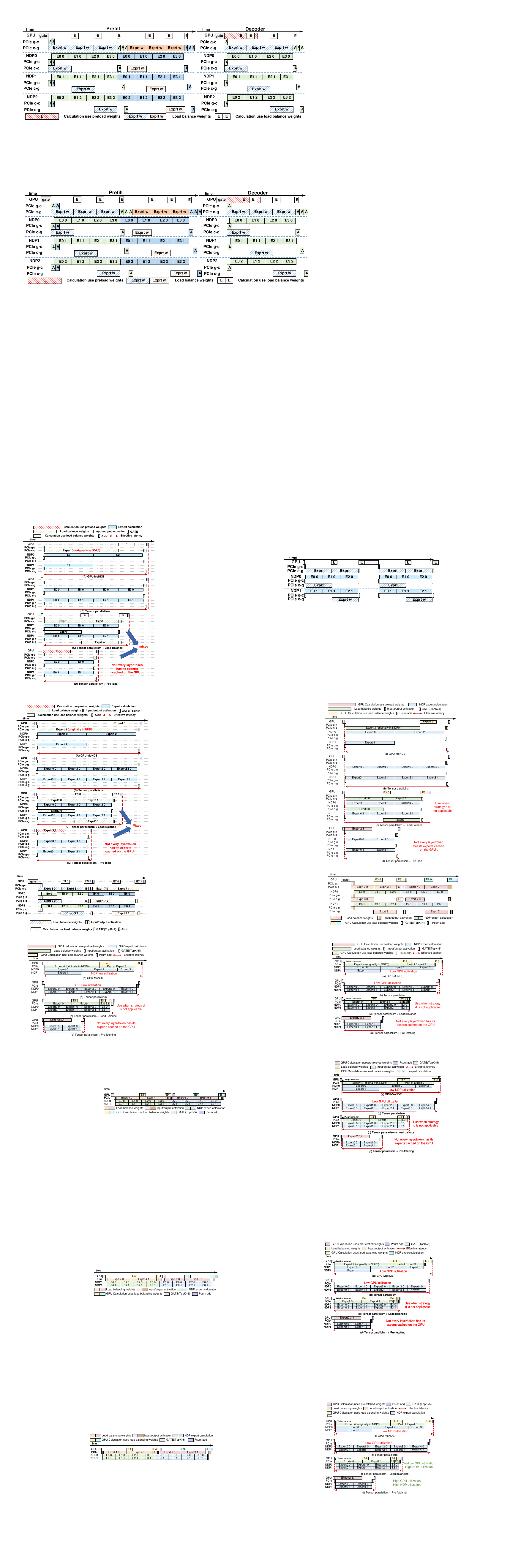}
    \vspace{-18pt}
    \caption{\qititle{MoE workflow scheduling in the prefill stage within our framework.}}
    \label{fig:prefill_strategy}
    \qicamera{\vspace{-17pt}}
\end{figure}

\chao{To mitigate this inefficiency, we propose implementing tensor parallelism across NDPs during the MoE computation stage.}
\qi{This strategy has historically been overlooked in the MoE scheduling research, largely due to the perceived high communication overhead of tensor parallelism. In single-batch, multi-NDP settings, however, the load imbalance introduced by expert parallelism outweighs the communication cost of tensor parallelism.}
As depicted in Fig.~\ref{fig:decode_strategy}b, tensor parallelism uniformly 
\qi{partitions each expert's computation across all available NDPs. Consequently, regardless of which expert is activated, each NDP is assigned a portion of the computation.}
\chao{Fig.~\ref{fig:tp} illustrates our tensor parallelism strategy employed in MoE layer computations, using three NDP devices as an example.}
\chao{We utilize a two-stage partitioning approach.}
\chao{The GEMV operations in the first two linear layers are partitioned by columns, while that in the last linear layers adopt row-wise partitioning.}
\chao{This approach efficiently distributes computational workload and memory requirements across devices to generate the final partial sum.}
\chao{However, while tensor parallelism achieves high utilization of NDPs in Fig.~\ref{fig:decode_strategy}b, the GPU remains idle during the decode stage, leaving room for further optimization. This observation motivates the need for load balancing between GPU and NDP to fully leverage both computational resources.}

\subsection{Load Balancing} \label{sec:lb}
\chao{To utilize idle GPU resources during NDP computation, we propose the load balancing strategy shown in Fig.~\ref{fig:decode_strategy}c and Fig.~\ref{fig:prefill_strategy}. Existing strategies like MoNDE\cite{kim2024monde} suffer from two limitations: they ignore the imbalanced workload distribution among activated experts and oversimplify balance conditions by only equating NDP computation time with weight transmission time, neglecting}
\qi{differences in the number of input tokens between the prefill and decode stages. Furthermore, they use a scaling factor based on historical data to correct the load balancing equation.}


\chao{We reformulate the balance condition based on tensor parallelism.
The key insight is 
\qi{that the $E_g$ value, \zong{denoting the number of experts assigned to the GPU},
should be determined to balance the total time spent on weight transmission and GPU computation against the total time for NDP calculation and input/output activation data transmission, rather than requiring adjustment via a periodic scaling factor as in MoNDE \cite{kim2024monde}}.
We also define $T_g$ as GPU expert computation latency, $T_n$ as NDP-DIMM expert computation latency after tensor parallelism, $T_w$ as expert transmission time, $T_a$ as activation transmission time.
$N$ represents the number of enabled NDP-DIMMs, and $S$ represents the sequence length.
\qicamera{Since the last transmission cannot be hidden, $S-1$ is actually used.}
The workload parameters include $\text{TopK}$ as the number of activated experts per layer, $E_n$ as the number of experts on NDP-DIMM, where $E_n + E_g = \text{TopK}$.}
\qi{Note that most computations on the GPU are overlapped by weight transfer; therefore, $E_{g'}$ denotes the unhidden computing time. 
For analytical tractability, $E_{g'}$ is defined as one N-th of the fractional part of $E_g$, reflecting the fraction of GPU expert computation that is not overlapped by N-way weight transfer; when $E_g$ is an integer multiple of $N$, a nominal value of $1/N$ is used instead.}

\textbf{Load balancing condition in decode stage:}
\begin{equation}
\label{eq:1}
T_w \cdot E_g + T_g \cdot E_{g'} = (N + 1) \cdot T_a + T_n \cdot E_n.
\end{equation}

\textbf{Load balancing condition in prefill stage:}
\begin{equation}
\label{eq:2}
T_w \cdot E_g + T_g \cdot E_{g'} + (S-1) \cdot T_a \cdot N = (N + 1) \cdot T_a + T_n \cdot E_n.
\end{equation}


\chao{The proposed load balancing conditions are formulated differently for decode and prefill stages due to their distinct computational characteristics. 
\zong{Equation~\eqref{eq:1}}
defines the balance condition for the decode stage, while 
\zong{Equation~\eqref{eq:2}}
accounts for sequence length dependencies in the prefill stage. \qi{Using these equations, the balanced expert $E_g$ can be calculated.} 
\qi{As illustrated in Fig.~\ref{fig:decode_strategy}c, load balancing enables a further improvement in the acceleration of the MoE stage.}
While this approach optimizes expert allocation, the frequent weight transfers required by dynamic load balancing may be limited by PCIe bandwidth, prompting us to further optimize runtime transmission overhead.}

\begin{figure} [t]
    \centering
    \includegraphics[width=1\linewidth]{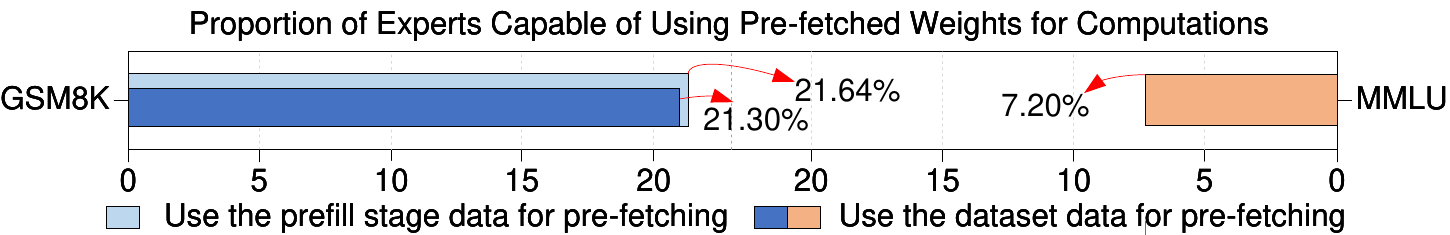}
    \vspace{-18pt}
    \caption{\qicamera{Utilization of prefetched weights in Qwen3-30B-A3B \cite{yang2025qwen3} is compared between analyses derived from the full calibration datasets (GSM8K, MMLU) and from dataset-free prefill-stage data. The comparison is conducted using 20 randomly sampled GSM8K inputs.}}
    \label{fig:dataset_specific}
    \qicamera{\vspace{-13pt}}
\end{figure}

\begin{figure} [t]
    \centering
    \includegraphics[width=1\linewidth]{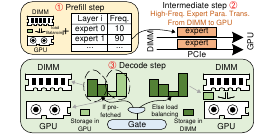}
    \vspace{-18pt}
    \caption{\qititle{Dataset-free pre-fetching strategy. It is divided into three steps: \ding{172}Prefill step, \ding{173}Intermediate step, and \ding{174}Decode step.}}
    \label{fig:prefetch}
    \qicamera{\vspace{-10pt}}
\end{figure}

\subsection{Dataset-free Pre-fetching} \label{sec:pre-fetch}
\chao{Pre-fetching expert weights into GPU memory before they are needed can significantly reduce PCIe transfer latency and improve GPU utilization}
Conventional pre-fetching strategies\cite{fang2025klotski}, however, typically rely on a pre-profiling of a calibration dataset to ascertain the activation patterns of experts,
\qi{incurring hundreds to thousands of extra inference passes and leading to dataset-dependent variability in the utilization of pre-fetched weights, as shown in Fig.~\ref{fig:dataset_specific}.
Although \cite{zhang2025daop} mentions that active experts have similarities in the prefill and decode stages, it still used the calibration dataset for initialization and introduced accuracy errors.}
\chao{Our strategy eliminates this limitation through a three-step pipeline as shown in Fig.~\ref{fig:prefetch}, which dynamically learns expert activation patterns during actual inference execution without calibration data.}
It is divided into three steps: \ding{172}prefill, \ding{173}intermediate, and \ding{174}decode.
\chao{The core innovation\chaoRtwo{, as illustrated in Fig.~\ref{fig:dataset_specific},} leverages activation patterns observed in the prefill stage to guide pre-fetching decisions for the decode stage, transforming costly offline profiling into lightweight runtime learning that adapts to actual workload characteristics.}

\qi{\textbf{\ding{172}Prefill step: dynamic information collection.}} 
\chao{As shown in Fig.~\ref{fig:prefetch}\ding{172}, during the prefill step, the system processes the input prompt while simultaneously collecting expert activation statistics in real-time.}
\chao{As FFN computations execute using the load balancing strategy from \zong{Equation~\eqref{eq:2}}, the system continuously monitors which expert modules are activated and tracks their frequency patterns.}
\qi{This dynamically collected activation data forms an instantaneous, task-specific expert table \chao{without requiring offline calibration datasets}, which serves as a direct basis for pre-fetching decisions in the subsequent step.}
\qi{Its pre-fetched expert utilization rate is even higher than that achieved with calibration datasets, as shown in Fig.~\ref{fig:dataset_specific}.}

\qi{\textbf{\ding{173}Intermediate step: weight pre-fetching.}
\chao{Based on the activation patterns identified during the prefill step, this intermediate step selectively pre-fetches the most frequently activated experts into GPU memory, \qi{as depicted in Fig.~\ref{fig:prefetch}\ding{173}.}}
\chao{\qi{The number of experts to be pre-fetched is determined by available GPU memory to ensure optimal resource utilization: at each layer, $x$ experts are pre-fetched, with $x$ initialized at 1 and incremented until no additional experts can be accommodated by the GPU.}
\qi{Since a single, small data transfer relative to the overall inference workload is performed during expert pre-fetching, minimal overhead is incurred.
In parallel, 
\zong{the remaining critical experts} are staged for immediate access during decode step.}}}

\qi{\textbf{\ding{174}Decode step: \chao{adaptive computation execution.}}
\chaoRtwo{As illustrated in Fig.\ref{fig:prefetch}\ding{174}, this step operates as follows: if the activated experts have been pre-fetched to the GPU, computation proceeds on the GPU; otherwise, non-prefetched experts are executed within the NDPs, as shown in Fig.\ref{fig:decode_strategy}d.}
\chaoRtwo{However, maximizing the number of experts computing on the GPU does not necessarily yield optimal performance.}
\chaoRtwo{A balance constraint must be maintained where GPU compute time should equal the combined NDP compute and transfer times.}
According to \zong{Equation~(\ref{eq:emax})}, $E_{max}$ can be estimated, which is the maximum number of experts that can be computed on the GPU.
In addition, when none of the activated experts are pre-fetched, the framework seamlessly switches to the standard load balancing strategy to handle the computing requests, as presented in Fig.~\ref{fig:decode_strategy}c. This fault-tolerant backup design ensures the maximum utilization of GPUs and NDP-DIMMs.}

\qi{\textbf{Maximum schedulable expert parameters to GPU:}}
\begin{equation}
\label{eq:emax}
T_g \cdot E_{max} = (\text{TopK} - E_{max}) \cdot T_n + (1 + N) \cdot T_a.
\end{equation}

%% file: Sections/4-experimental.tex
\begin{table}[tb]
\centering
\caption{GPU-NDP System Configuration}
\noindent
\begin{tabular}{c|c|c|c} 
\toprule 
\multicolumn{2}{c}{\textbf{GPU configuration}} \vline & \multicolumn{2}{c}{\textbf{NDP-DIMM configuration}} \\ 
\midrule 
Frequency & 2.30 GHz & DDR4 type & 3200 MT/s, 32GB \\ 
GDDR & 16GB GDDR7 & Ranks & 4 / DIMM \\
SM Count & 84 & Bankgroups & 8 / Rank \\
Tensor Cores & 336 & Banks & 4 / Bankgroup \\
Interface & PCIe 5.0 & Multipliers & 64 / DIMM \\
\bottomrule
\end{tabular}
\label{tab:dimm}
\qicamera{\vspace{-10pt}}
\end{table}

\section{Experimental Results} \label{sec:experimental}




\subsection{Experimental Setup}

\subsubsection{\textbf{System}}

\chao{To evaluate our proposed scheduling framework, we employ a GPU-NDP system architecture as shown in Fig.~\ref{fig:system}.}
\chao{The system comprises a consumer-grade NVIDIA RTX 5080 GPU~\cite{nvidiaRTX5080} with 16GB GDDR memory and PCIe 5.0 x16, paired with an Intel i7-14700 processor~\cite{Intel14700} that provides up to 192GB of system memory with a maximum bandwidth of 89.6 GB/s.}
\chao{The GPU-NDP system supports up to six 32GB NDP-DIMMs, each offering an internal bandwidth of 102.4GB/s as detailed in Table~\ref{tab:dimm}.}
\chao{For performance evaluation, we utilize AttAcc!~\cite{park2024attacc} to simulate GPU performance characteristics and employ a modified version of Ramulator 2.0~\cite{luo2023ramulator} to} evaluate the performance of the DIMM devices with computing capabilities,
\qicamera{where the simulation tool's accuracy has been verified for GPU and PIM operations in~\cite{luo2023ramulator}.}


\begin{table}[tbp]
\centering
\caption{Workload Characteristics of MoE-Based Models}
\resizebox{0.48\textwidth}{!}{  
\begin{tabular}{c|c|c|c|c|c|c}
\toprule
\textbf{Model} & \textbf{E Para.} & \textbf{\#E} & \textbf{TopK} & \textbf{Hidden} & \textbf{Interm.} & \textbf{\#Layer} \\
\midrule
DeepSeek-MoE\cite{dai2024deepseekmoe} & 15.4B & 64 & 6/2$^\vartriangle$ & 2048 & 1408 & 27/28$^*$ \\
Qwen3-30B-A3B\cite{yang2025qwen3} & 29.0B & 128 & 8 & 2048 & 768 & 48 \\
Phi-3.5-MoE\cite{haider2024phi} &40.3B & 16 & 2 & 6400 & 4096 & 32 \\
Mixtral-8x7B\cite{jiang2024mixtral} & 42.0B & 8 & 2 & 4096 & 14336 & 32 \\
\bottomrule
\end{tabular}
}
  \begin{tablenotes} 
    \item $^\vartriangle$6/2 indicates the activation of 6 routing experts and 2 shared experts. 
    \item $^*$27/28 indicates that 27 out of 28 layers are MoE layers. 
 \end{tablenotes} 
\label{tab:Workload}
\qicamera{\vspace{-10pt}}
\end{table}

\begin{figure} [t]
    \centering
    \includegraphics[width=1\linewidth]{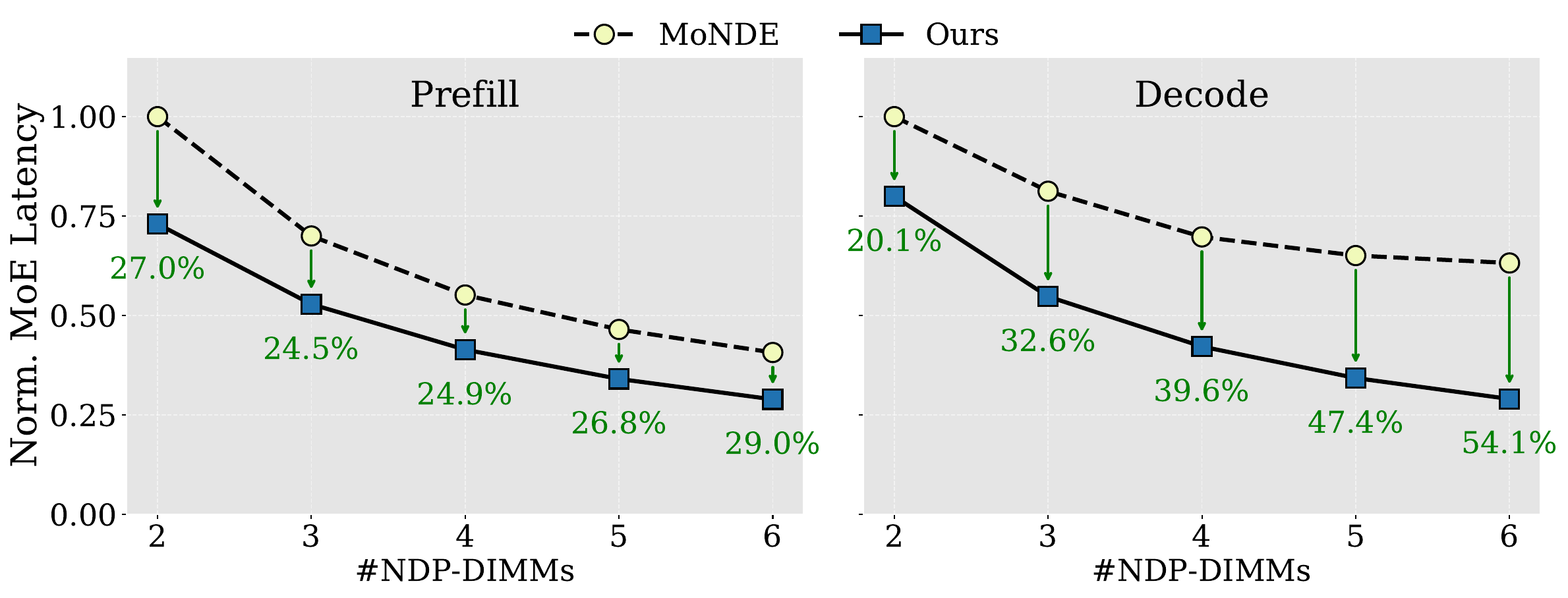}
    \vspace{-18pt}
    \caption{The normalized MoE latency of the prefill and the decode stage in comparison with SOTA MoNDE~\cite{kim2024monde}.}
    \label{fig:prefill_decode}
    \qicamera{\vspace{-10pt}}
\end{figure}

\begin{figure} [t]
    \centering
    \includegraphics[width=1\linewidth]{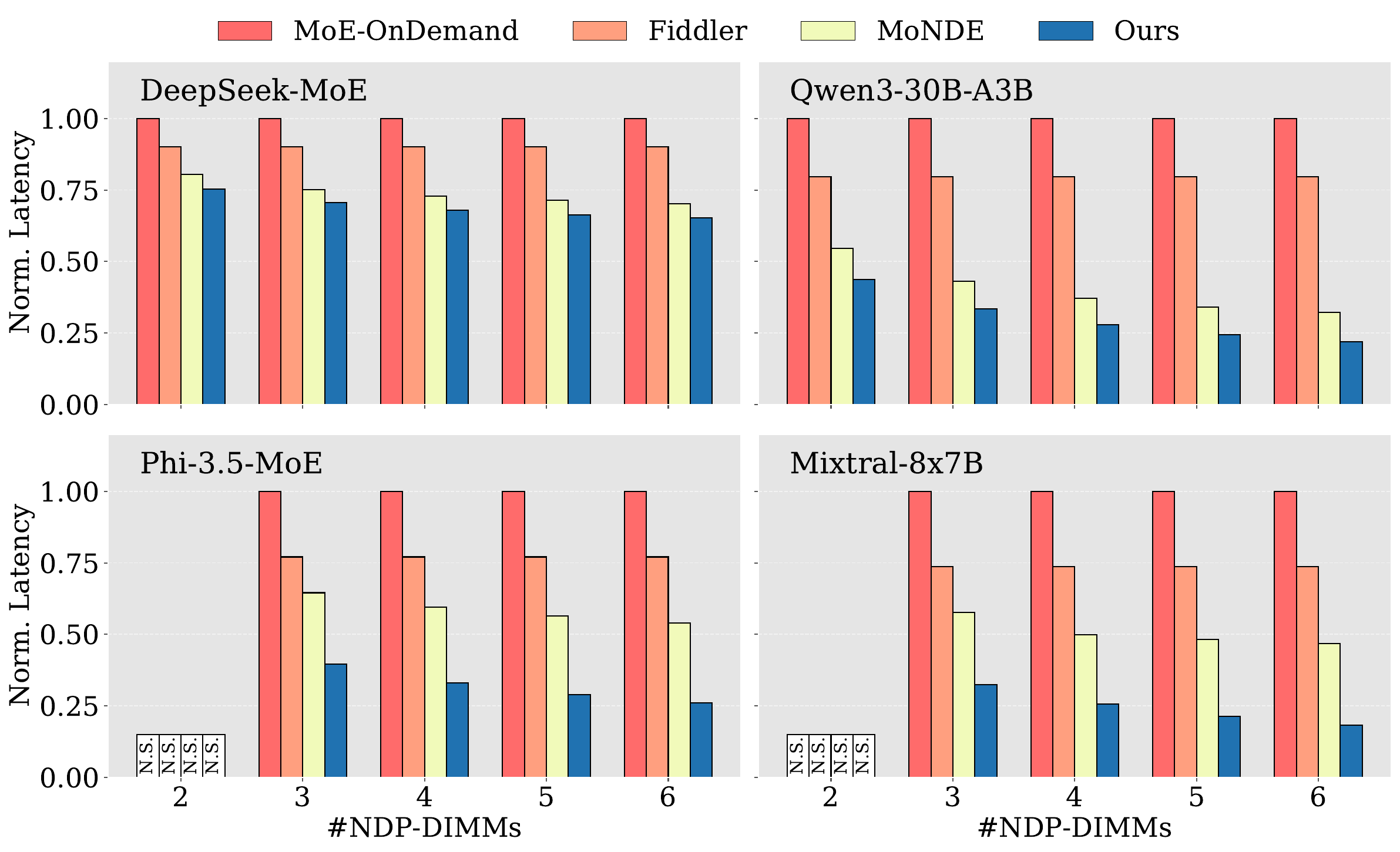}
    \vspace{-18pt}
    \caption{End-to-end model performance with varying number of NDP-DIMMs normalized to MoE-OnDemand~\cite{aminabadi2022deepspeed}. N.S. indicates "Not supported", meaning that this configuration is unable to accommodate all the MoE parameters.}
    \label{fig:end_to_end}
    \qicamera{\vspace{-10pt}}
\end{figure}

\begin{figure} [t]
    \centering
    \includegraphics[width=1\linewidth]{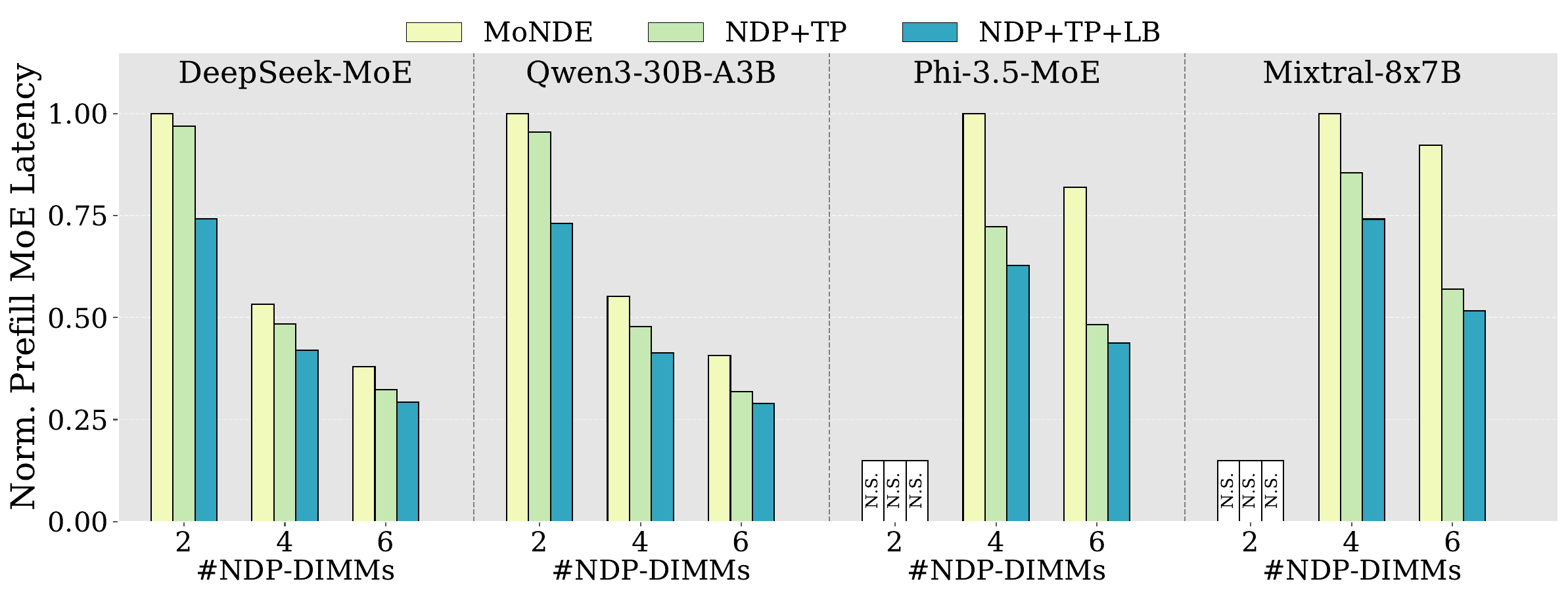}
    \vspace{-18pt}
    \caption{The impact of tensor parallelism and load balancing on MoE computation in the prefill stage. N.S. indicates "Not supported", meaning that it is unable to accommodate all the MoE parameters.}
    \label{fig:ablation_study_prefill}
    \qicamera{\vspace{-10pt}}
\end{figure}

\subsubsection{\textbf{Models}}

\chao{We use 4 representative MoE models from the HuggingFace repository for evaluation: DeepSeek-MoE~\cite{dai2024deepseekmoe}, Qwen3-30B-A3B~\cite{yang2025qwen3}, Phi-3.5-MoE~\cite{haider2024phi}, Mixtral-8x7B~\cite{jiang2024mixtral}.}
\chao{These models exhibit diverse architectural characteristics, as summarized in Table~\ref{tab:Workload}.}
\chao{The number of activated experts per layer ranges from 2 to 8, and the total number of experts per layer varies from 8 to 128.}
\chao{Notably, DeepSeek-MoE~\cite{dai2024deepseekmoe} incorporates shared experts that are utilized by all tokens, distinguishing it from the other three models.}
\chao{Given our focus on local deployment scenarios for personal devices, all experiments are conducted with a batch size of 1. 
The input and output sequence lengths are uniformly set to 512 tokens across all experiments to ensure consistent evaluation conditions.}



\subsubsection{\textbf{Baseline}}

\chao{To provide comprehensive performance comparisons, we evaluate against 3 types of baseline approaches.}
\chao{1) MoE-OnDemand~\cite{aminabadi2022deepspeed}: A GPU on-demand expert weight transmission system where MoE parameters are offloaded to system memory and activated experts are dynamically loaded to the GPU during runtime.}
\chao{2) Fiddler~\cite{kamahori2025fiddler}: A SOTA GPU-CPU heterogeneous system that offloads MoE parameters to memory but performs expert computations on the CPU rather than the GPU.}
\chao{3) MoNDE~\cite{kim2024monde}: A SOTA NDP system that stores and computes MoE components using dedicated processing units, transmitting activation values rather than model weights to reduce data movement overhead.}
\qicamera{The deployment context aligns with the edge‑centric scenario examined in this work.}


\subsection{\chao{Evaluation of Isolated Prefill and Decode Performance}}

\chao{We analyze the MoE prefill and decode performance by comparing our framework against MoNDE~\cite{kim2024monde}.}
\chao{Fig.~\ref{fig:prefill_decode} presents the expert computing latency between MoNDE and our approach using Qwen3-30B-A3B as an example.}
\chao{The results demonstrate significant performance improvements, achieving an average speedup of 1.36$\times$ in the prefill stage and 1.69$\times$ in the decode stage, respectively.}
\chao{This performance gain stems from our more efficient hybrid scheduling strategy. While MoNDE distributes expert workloads cyclically based on arithmetic intensity, it fails to achieve optimal utilization across all NDP units.}
\chao{In contrast, our hybrid method achieves more balanced resource utilization across different scenarios.}
\chao{Examining the two stages in detail, the decode stage shows particularly pronounced gains compared with the prefill stage.}
\chao{In the prefill stage, although concurrent multi-token processing under the MoNDE scheme yields a relatively uniform distribution of activated experts across NDP units, greater speedup is nevertheless achieved by our framework through even more balanced participation of NDPs.}
\chao{In the decode stage, where a single token is generated serially, significant load imbalance between NDPs is observed in the MoNDE scheme. In contrast, our approach maintains the same load across all NDPs, leading to higher efficiency.}


\subsection{Evaluation of End-to-End Performance}
We then evaluate the end-to-end performance of the proposed framework \chao{with varying numbers of NDP-DIMMs}. 
\qi{Fig.~\ref{fig:end_to_end} presents the end-to-end latency results for four MoE models, normalized against MoE-OnDemand performance.}
\chao{Our framework demonstrates significant performance improvements across all baseline methods. Specifically, we achieve a maximum speedup of 5.49$\times$ and 4.05$\times$ compared to the parameter offloading approaches MoE-OnDemand~\cite{aminabadi2022deepspeed} and Fiddler~\cite{kamahori2025fiddler}, respectively, and 2.56$\times$ over the NDP baseline MoNDE~\cite{kim2024monde}.}
\chao{Since our acceleration mainly targets MoE computations,} the larger the number of parameters in the MoE layer, the more significant the acceleration \chao{our framework achieves}. 
\chao{The effectiveness of our approach is attributed to the integration of tensor parallelism, load balancing, and pre-fetching techniques, which collectively enable efficient expert distribution across NDP-DIMMs while optimizing both NDP utilization and GPU resource allocation.}

\subsection{Ablation Study for Our Proposed Techniques}

\chao{An ablation study is conducted to quantify the individual and joint contributions of our three key techniques: tensor parallelism, load balancing, and pre-fetching.}
\chao{We compare different combinations of these techniques against the NDP baseline, i.e, MoNDE~\cite{kim2024monde}: tensor parallelism alone (NDP+TP), tensor parallelism with load balancing (NDP+TP+LB), tensor parallelism with pre-fetching (NDP+TP+PRE), and the full combination (NDP+TP+LB+PRE).}
\chao{For all ablation configurations, we pre-fetch the \qicamera{most probable} TopK experts per layer to GPU memory after the prefill stage for each evaluated model. However, due to GPU memory constraints, Mixtral-8x7B~\cite{jiang2024mixtral} is limited to pre-fetching only one expert per layer.}

\begin{figure} [t]
    \centering
    \includegraphics[width=1\linewidth]{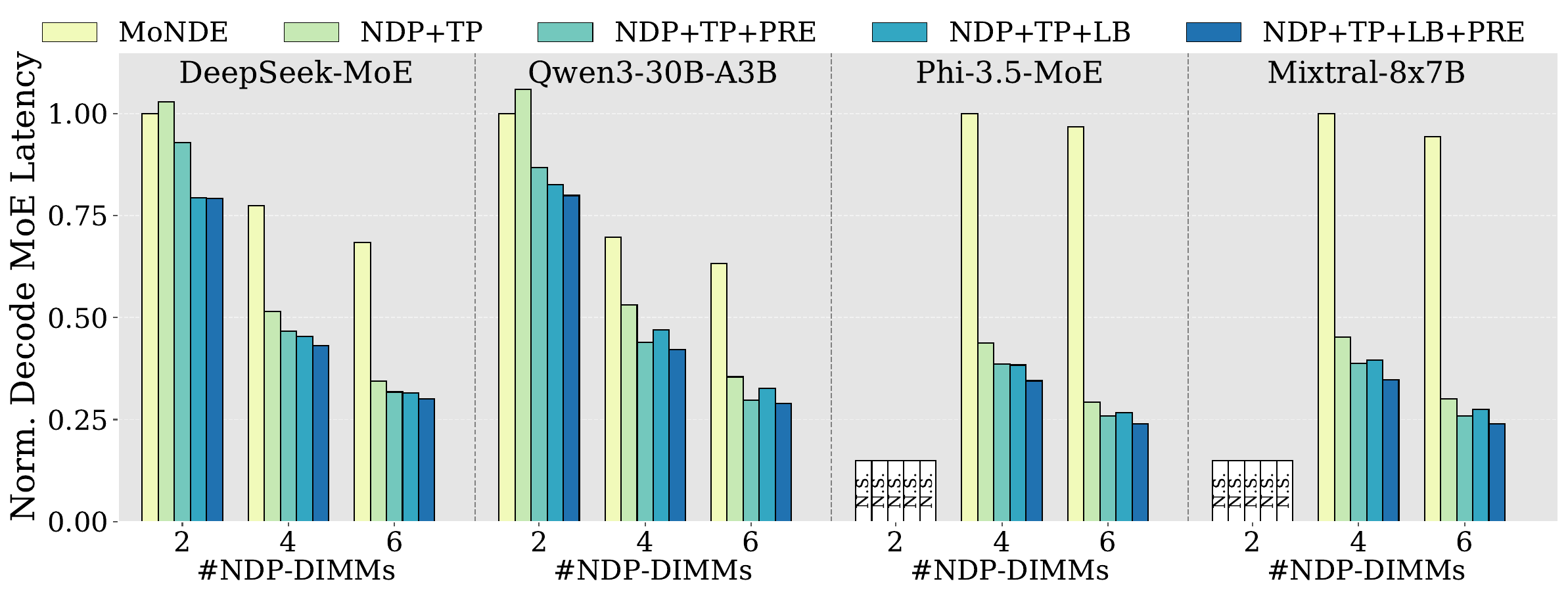}
    \vspace{-18pt}
    \caption{The impact of tensor parallelism, load balancing, and dataset-free pre-fetching on MoE computation in the decode stage. N.S. indicates "Not supported", meaning that it is unable to accommodate all the MoE parameters.}
    \label{fig:ablation_study_decode}
    \qicamera{\vspace{-10pt}}
\end{figure}

\qi{As presented in Fig.~\ref{fig:ablation_study_prefill}, during the prefill MoE stage, \chao{the NDP+TP and NDP+TP+LB strategies achieve average speedups of 1.26$\times$ and 1.46$\times$ over MoNDE, respectively.}}
\qi{Since the pre-fetched experts are only applicable during the decode MoE stage, the PRE-related strategies are not shown in Fig.~\ref{fig:ablation_study_prefill}.}
\qi{In the decode MoE stage, as depicted in Fig.~\ref{fig:ablation_study_decode}, an average speedup of 1.99$\times$ is achieved with NDP+TP alone, while average speedups of 2.19$\times$ and 2.23$\times$ are achieved with NDP+TP+LD and NDP+TP+PRE, respectively, when compared with MoNDE. 
\chao{The combination of all three strategies (NDP+TP+LB+PRE) consistently yields the highest average speedup of 2.41$\times$ across all evaluated models.}}

\chao{As shown in Fig.~\ref{fig:ablation_study_decode}, two notable observations emerge from the decode MoE stage results.}
\chao{First, when using 2 NDP-DIMMs, the NDP+TP method performs slightly worse than MoNDE for DeepSeek-MoE and QWen3-30B-A3B models. Individual NDP units have much lower compute capability than a GPU, making tensor parallelism less effective when NDP resources are limited, while MoNDE can leverage both GPU and NDP coordination.}
\chao{Second, the benefit of hybrid strategies over tensor parallelism alone decreases as we add more NDP-DIMMs.}
\chao{With more NDP-DIMMs, expert computation becomes much faster, shortening the time window for parameter transfer and GPU computation, which reduces the relative gains from load balancing and pre-fetching optimizations.}

%% file: Sections/5-conclusion.tex
\section{Conclusion} \label{sec:conclusion}

\chao{This paper presents a scheduling framework that leverages GPU-NDP systems to accelerate MoE model inference in edge deployment scenarios. The proposed framework exploits tensor parallelism to partition expert parameters across multiple NDP units, implements a load-balancing-aware scheduling algorithm to optimize resource utilization across both NDP units and GPU, and employs a dataset-free pre-fetching strategy to proactively load frequently accessed experts. Experimental results illustrate that with these collaborative innovations, our framework achieves \qicamera{2.41$\times$ on average and} up to 2.56$\times$ reduction in end-to-end latency compared to the state-of-the-art NDP-based MoNDE approach, significantly enhancing MoE inference efficiency in resource-constrained edge environments.}